\documentclass[smallextended]{svjour3}       
\RequirePackage{fix-cm}
\smartqed  
\usepackage{graphicx}
\usepackage{subcaption}
\usepackage{pdfpages}
\usepackage{longtable}
\usepackage{natbib}
\usepackage{rotating}
\usepackage{hyperref}
\usepackage{endfloat}

\usepackage{atbegshi}
\def\makeheadbox{\relax}
\begin{document}
\journalname{}

\title{Evacuation decisions in response to natural disasters:  Insights from a large-scale social media survey}


\titlerunning{Surveying Displaced Populations with Facebook}        

\author{
    Paige Maas \and Zack Almquist \and Eugenia Giraudy \and JW Schneider
}


\institute{
           Z. Almquist \at
Seattle, WA \\ 
          \email{zalmquist@uw.edu}
}

\date{\today}

\maketitle

\begin{abstract}

Evacuation in response to natural disasters is a complex process involving multiple decision-makers at the personal, household, community, and government levels.  Consequently, many disparate factors influence who evacuates, when, and how to respond to a nearby disaster.  In this paper, we leverage a novel method of data collection through social media to explore the evacuation response decisions of people in areas affected by the 2019-2020 Australian bushfires.  We explore the validity of this data collection method for generating plausible estimates of evacuation and its ability to supplement cell phone location data using survey responses.  Ultimately, we identify several key factors influencing household decisions on evacuation, specifically focusing on the phenomenon of household members evacuating or returning from evacuation at different times. 

\end{abstract}

\clearpage

\section{Introduction}
\label{intro}

The evacuation and displacement of households is a typical consequence of natural disasters.  In some cases, these evacuations are an immediate result of the physical destruction of a dwelling.  However, evacuations, in general, are a response to safety risks associated with any given disaster. Though governments often mandate evacuations in response to disasters or disaster risks, enforcement of such mandates is rarely binding or enforced.  Indeed, evacuation at some level does require the active decision-making or cooperation of affected or at-risk persons or households.  Understanding factors associated with personal or household decisions to evacuate and when is critical to mitigating the potential effects of a disaster and fostering recovery in affected areas.

Understanding these motivations, however, is complicated by the difficulty associated with collecting information from affected persons.  Historically, large-scale post-disaster surveys are complicated by the difficulty of reaching people who have been displaced or evacuated following a disaster \citep{frankenberg2014demography}. Recently, new data sources and methods have become available to measure the population properties of evacuation and displacement behavior due to natural disasters; much of this work has focused on social media, and modern geo-location data are now able to assist researchers in making inferences on this population in a cost-effective manner. This line of inquiry includes administrative data (e.g. IRS data \citet{fussellRecoveryMigrationCity2014}; Consumer Panel Data \citet{dewaardOutmigrationReturnMigration2020}), geo-location data (e.g. Facebook Disaster Maps; \citet{maas2019facebook}), and social media data (e.g. Twitter; \citet{martinUsingGeotaggedTweets2020} or Facebook Marketing API data; \citet{alexander20}).

We extend this work by leveraging a social media platform (Facebook) to identify a population likely affected by the Australian bushfires of 2019-2020. 

From September 2019 to March 2020, Australia experienced one of the worst bushfires in modern history \citep{kganyago2020assessment}. Over this period, it is estimated that 186,000 square kilometers were burned, and 5,900 buildings were destroyed. This included more than three thousand homes \citep{filkov2020impact} and displaced tens of thousands of Australians. To better understand the economic and demographic effects of this disaster in Australia, we surveyed Facebook users who could have been directly affected by the bushfires.\footnote{\citet{20192020AUSTRALIANBUSHFIRES} (IDMC) has attempted to estimate the economic and demographic effects of the Australian bushfires using the data in this article as well as other data sources.  Specifically, IDMC found that cost between \$44 million and \$52 million due to population displacement during and after the Bushfires.}  This, when combined with mobility estimates provided by the Data for Good program, could allow us to refine estimates derived from mobility data \citep{maas2019facebook} and to address hypotheses around how displacement affected people and households in the region -- information that cannot be obtained from observational migration data alone.

We explore methods for drawing displacement- and evacuation-related inferences by comparing a stratified sample of users who opt into providing their geo-location via mobile devices (a small subset of Facebook users) and a random sample of comparable users who are predicted to have been in the affected region before the disaster. We then use this frame to survey disaster response behaviors that supplement the limited information we get from the geo-location data.  By employing rejection sampling, we identify a final set of respondents who were actually in the disaster zone before the bushfire. In conducting this research, we partnered with Facebook's Data for Good program, which has already used mobility data to develop an approach that both identifies evacuation/displacement and relocation after disasters to provide real-time information about displacement to non-governmental organizations (NGOs) for disaster relief purposes \citep{maas2019facebook}.

This article is laid out in the following manner: (1) a brief overview of the social science of evacuation behavior to motivate the core set of research questions; (2) a review of the innovations related to our methodology; (3) a review the survey sample and comparison to the general Australian Facebook population and the general Australian population; (4) an analysis of the survey associated with our research questions (5) a discussion of policy implications and areas for future research.

\section{Social science of short-term displacement and evacuation behavior}


There is a robust literature on \emph{who evacuates} and \emph{why} in the social sciences in response to natural disasters. For a complete review, see \citet{thompson2017evacuation}. Prior work tends to focus on \emph{predicting evacuation behavior}  either after a natural disaster has occurred (e.g., fire or flood) or in the period before a natural disaster  \cite[for example][]{adeolaHurricaneKatrinalinkedEnvironmental2017,aguirreEvacuationCancunHurricane1991,bakerGeographicalVariationsHurricane1979,bakerPredictingResponseHurricane1979,batemanGenderEvacuationCloser2002,brackenridgeDimensionsHumanAnimal2012,bukvicAttitudesRelocationFollowing2017,dowEmergingHurricaneEvacuation2002,hasanBehavioralModelUnderstand2011,hasanTransferabilityHurricaneEvacuation2012,horneyFactorsAssociatedEvacuation2010,loweTrajectoriesPsychologicalDistress2013,mesa-arangoHouseholdlevelModelHurricane2013a,mesa-arangoHouseholdlevelModelHurricane2013,morssUnderstandingPublicHurricane2016,murray-tuiteChangesEvacuationDecisions2012,phamEvacuationDepartureTiming2020,sadriAnalysisHurricaneEvacuee2014,thiedeHurricaneKatrinaWho2013}. 

The literature has explored in depth many factors influencing evacuation and displacement decisions. Social networks, \citep{adeolaKatrinaCataclysmDoes2009}, concern for the safety of one's belongings \citet{aguirreEvacuationCancunHurricane1991}, and housing tenure \citep{bakerPredictingResponseHurricane1979} have all been demonstrated to be relevant.  

The findings on demographic variables such as gender are more mixed.  At least one prior study has found no real association between evacuation decisions and demographic variables such as age and gender \citep{bakerPredictingResponseHurricane1979}. Others found that women were more likely to have an evacuation plan, but men were more likely to evacuate  \citep{batemanGenderEvacuationCloser2002}. 

Research focusing on wildfires is much more limited, but there is an emerging literature about evacuation during these types of disasters. Work has found that those who evacuated had more interest and access to information on evacuation, road closures, and home protection for wildfires compared to those who did not evacuate \citep{mccaffreyDifferenceInformationNeeds2013}. However, it has been shown that, in general, people in wildfire situations have a strong desire to shelter in place \citep{covaProtectiveActionsWildfires2009}. Recent work on Australian Bushfires observes gender differences in response \citep{tylerBushfiresAreMen2013}. It explores the relationship between masculinity and fire preparedness, noting that men are over-represented in death statistics for wildfires \cite{covaProtectiveActionsWildfires2009}.  It has been shown that households who have evacuated have a plan to do so and that many of those who stayed had a plan to protect their property, which is suggestive of economic dimensions to evacuation decisions \citep{mclennanHouseholdersSafetyrelatedDecisions2013}.

\citet{hasanBehavioralModelUnderstand2011} reviews the major literature on \emph{warnings}, \emph{risk} and finally \emph{observed behavior} in response to natural disasters and evacuation. This includes a basic overview of who evacuates and who does not, including attempts to model evacuation compliance \citep{bakerPredictingResponseHurricane1979}. Of special relevance to this paper is the work of \citet{dashEvacuationDecisionMaking2007}, who studied household decision-making in response to natural disasters. Others have demonstrated that households utilize information from local authorities, peers, media, and geography.  We also explore these decisions with other factors to provide a detailed look at the displacement parameters \citep{dashEvacuationDecisionMaking2007}.

\section{Data sources for evacuation and displacement due to natural disasters}

Natural disasters are, by definition, sudden and hard to survey prospectively. Due to the nature of displacement or evacuation, it can be difficult to survey those affected by natural disasters.  Finding a suitable sampling frame is a major issue. It has received attention from the larger research community (see \citet{thompsonEvacuationNaturalDisasters2017}, which reviews major data collection activities pre and post-disaster over the last 40 years). In practice, most research of displaced or evacuated behavior occurs around 6 months after the disaster \citep{thompsonEvacuationNaturalDisasters2017}. Online data collection is still quite rare; for example, in \citet{thompsonEvacuationNaturalDisasters2017}'s review, only one data set was collected online, while the rest were collected via mail, telephone, or in person with the sampling frame being either convenience sample, phone sample (random digit dialing), or household sample. The obvious limitation of such frames is the inability to locate people who have not returned or whose homes or phones are still inaccessible. Furthermore, declining response rates reduce the effectiveness of random digit dialing methods.  To overcome these limitations, researchers have employed administrative data and various online data sources such as Facebook ad systems, Twitter, and cell phone or app geolocation data, which we discuss below.

\subsection{Classic methods of interviews and surveys}

\subsection{Administrative data}

Following longer-term displacement and evacuation is work centered around using administrative data, such as the Internal Revenue Service (IRS) county migration data \citep{fussellRecoveryMigrationCity2014}, and for shorter-term estimates, the Federal Reserve Bank of New York/Equifax Consumer Credit Panel (CCP) \citep{dewaardOutmigrationReturnMigration2020}. The IRS data has been used to look at the 3-4 year recovery period post Hurricane Katrina for return migration \citep{fussellRecoveryMigrationCity2014}, and the CCP data for continued estimation of displacement due to Hurricane Maria two years after \citep{dewaardOutmigrationReturnMigration2020}. Data of this nature tends to be limited in time (e.g., the IRS data is only available yearly) or income (e.g., CCP requires the user to have a credit card).

\subsection{Social Media}

\subsubsection{Twitter}

Twitter, a popular social media application, has been used to estimate evacuation and displacement of people post Hurricane Maria in Puerto Rico \citet{martinBridgingTwitterSurvey2020}. For example, \citet{martinBridgingTwitterSurvey2020} demonstrated that geotagging tweets provides a representative sample of individuals aged 18–54 years. This method was a good complement to an elderly-biased questionnaire survey. This work demonstrated that Twitter provided a good source of information on the timing and destination of displaced persons and whether they returned to their homes. Following Hurricane Maria, it was estimated that in the Twitter sample, 8.3\% of residents relocated, and nearly 4\% continued to be displaced 9 months later  \citet{martinUsingGeotaggedTweets2020}. Other work has used Twitter to characterize aspects of displacement such as destination, relative impact, and where the disaster is most severe \citep{yumMiningTwitterData2020}. 

\subsubsection{Facebook}

Facebook has been used to address immigrant assimilation \citep{stewart2019rock}, world fertility \citep{ribeiro2020biased}, and world migration stocks \citep{zagheni2017leveraging}. 
Recently,  Facebook's Marketing API has been used to estimate displacement and evacuation from Hurricane Maria in Puerto Rico  \citep{alexander20}.\footnote{It is possible to argue this data is really administrative data, but as it comes from social media, we discuss it in this subsection.}  Working with the raw Facebook Marketing API data is problematic because it includes movement not related to an event in question, so \citep{alexander20} employ a Difference-in-Differences and data in the ACS to improve estimates of out-migration of Puerto Ricans due to hurricane Maria \citep{alexander20}. Further, the Facebook ads system has been to recruit survey participants to answer issues of mobility and geographic partitioning \cite{blondel2015survey}.

Facebook's advertising platform has been shown to provide a good sampling frame for online surveys and helps with timeliness, coverage, and cost-effectiveness challenges.
\citep{grow2020addressing}. While Facebook's user base is a rough cross-section of the overall population with internet access, \cite{grow2020addressing} advises stratifying by demographic characteristics, which are known to correlate with the outcome of interest and to post-stratify in the final analysis. There is a growing literature on how to re-adjust surveys administered on the Facebook ads \citep{zagheni2017leveraging} platform or survey system \citep{feehan2019using} for estimation of both online and offline populations.  For example, \\cite{schneider2019s} explores using Facebook-targeted advertisements to collect data on low-income shift workers in the United States.  Using social media, such as Facebook, as a survey platform for demographic work - especially for populations that are challenging or expensive to access - is increasingly of interest to the demographic community. We build on this work by introducing a rapid-response survey of post-disaster demographic and economic outcomes through the Facebook app.

\subsection{Geolocation data}

Using cell phone and app-based data for social science is an active area of research.  This includes short-term displacement due to disasters. For example, \cite{lu2012predictability} used 1.9 million phone users' location data subsequent to, and for up to one year after, the 2010 Haitian earthquake and uncovered detailed and variable mobility patterns among the studied population. \cite{deville2014dynamic} follows up on this work and demonstrates the cost-effectiveness of using mobile phone network data for accurate and detailed maps of populations after a disaster. This is an active area of the literature with especially promising applications in estimating the population of countries that have historically had poor or non-existent demographic data sources \citep{tatem2017worldpop}.  For example, Facebook's Data for Good program is a broad initiative designed to provide data to humanitarian organizations to facilitate their important work in many fields, including disaster response and disease prevention. One such dataset is the Gender-Stratified Displacement Map (referred to hereafter as \emph{displacement maps}), which aims to quantify the magnitude of displacement following disasters and describe where the displaced or evacuated population has migrated.  This data enables the study of these population trends by gender. Facebook's Displacement Map dataset estimates how many people were displaced by a given disaster and where the population has shifted in the period following the event, aggregated at the city level. Specifically, the models identify a person as displaced if their typical nighttime location patterns are disrupted after the event compared to that person’s pre-disaster patterns.  These patterns are obtained from a user's Location History, an optional setting on the Facebook app that users can enable that provides precise locations \citep{HowFacebookLocation}.\footnote{Unfortunately for future research, Facebook no longer has location history data.}  Individual data is aggregated into a city-level transition matrix showing how many people are displaced or evacuated from one city to another for all source cities in the disaster-affected region for each day for a period following the event \citep{maas2019facebook}.

\section{Research questions on evacuation behavior, timing and demographics}

We proceed to explore the topic of the decision to evacuate at the individual and household levels in response to a natural disaster.  We suggest using a simple logistic regression on the decision to evacuate and whether to do so as a household or separately as an individual as a function of demographic characteristics.

\noindent
\textbf{Who evacuates?}

 We begin by identifying traditionally important questions from the literature about evacuation and displacement caused by disaster events. \citet{thompson2017evacuation} and others explore who evacuates and how evacuees and non-evacuees differ by demographic and economic factors.
 
We focus on who evacuated or displaced more than one night and who is still displaced 2 months after the disaster. Further, we are interested in the questions of differences by demographics and socio-economic variables. The appendix contains our survey instrument (Survey Appendix: Survey Questions and Logic), and in it, we ask respondents ``Did you leave your home for more than one night as a result of the bushfire?" with response options: Yes; No; I don't know. We also ask ``Did you return home at least one day before other members of your household?" with response options: Yes; No; I don't know (Survey Appendix: Survey Questions and Logic: 5 and 22). The survey follows with detailed questions about the length of time displacement, if an evacuation occurred before or after the disaster, what transportation was used, if the person returned home, and where they went while displaced. 

\noindent
\textbf{Who makes the decision to evacuate?}

 \citet{thompson2017evacuation} and others also document the importance of who decides to evacuate on whether or not a household does evacuate. We ask displaced persons how they decided to evacuate and explore how this differs by demographic characteristics. Specifically, Survey Appendix: Survey Questions and Logic: 19 asks ``Whose decision first led you to leave your home?" with response options: (1) My local, state, or national government; (2) A family member; (3) My own decision; (4) Someone else.

\noindent
\textbf{Where do people go when they evacuate?}

There is a long literature on the short-term and long-term displacement of people due to disasters \cite[e.g.][]{dynes1987sociology}. Similarly, we are interested in short-term displacement and long-term migration. For displaced persons, we ask where they went and how far away from the disaster event they traveled.  In the survey we ask (Survey Appendix: Survey Questions and Logic: 13) ``Where did you go when you left your home?" with response options: (1) Within the same city as my home; (2) A different city, but in \{Region\}\footnote{Where Region is replaced with the name of the local area, e.g. Adelaide Hills, South Australia.}; (3) A different state or territory in Australia; (4) A different country.

\noindent
\textbf{Under what circumstances do households split up during displacement?}

 \citet{thompson2017evacuation} and others also document the importance of household composition on displacement. In this work, we focus on household separation during a disaster. We seek to learn about this through two questions: Survey Appendix: Survey Questions and Logic: 10: ``When you were displaced for more than 3 nights did anyone in your household stay behind?" both with response options: Yes; No; I don't know. Survey Appendix: Survey Questions and Logic: 22: ``Did you return to your home at least one day before the other household members?"
 
\noindent
\textbf{Whose work is disrupted?}

A common question in disaster research relates to economic disruption or effects on people's ability to work \citep{rose2004defining}. We directly measure if people were unable to work and look at how demographics and socioeconomic status either increase or decrease one's likelihood of economic disruption. We assess this through Survey Appendix: Survey Questions and Logic: 35: ``Did leaving your home prevent you from working as much as you normally do?" with response options: Yes; No; Prefer not to say.

\section{Methodology}

One major issue with disaster research is establishing a sampling frame for the population of interest: all people in the region when a natural hazard event occurs. One solution to this problem is using social media platforms \citep{alexander20}. In this work, we use  Facebook’s survey infrastructure to obtain a sample of people in the region during the natural disaster. Surveying from the Facebook platform allows for rapid deployment of a survey post-disaster that is highly representative of Facebook users and, in the long term, could be employed to estimate offline populations \citep{feehan2019using}.  Facebook reported in January 2020 that it had 2.89 billion monthly active people and 2.26 billion daily active people on average \citep{fb2019}. In 2020, it was estimated that there were 4.93 Billion people online  \citep{worldStats2020} out of 7.70 Billion people \citep{UScensus}. 

\subsection{Australia Bushfires, 2019-2020}

The most recent census data for Australia places the country's total population at around 26 million \citep{auCensus} and estimates of the economic impact of bushfires ranging from 1.8 billion to 4.4 billion (AUD) \citep{OngoingImpactBushfires,butlerEconomicImpactAustralia2020}. Bushfires have become a major recurring natural disaster in Australia over that nation's history \citep{OngoingImpactBushfires}.

In this work, we focus on the \emph{Green Wattle Creek Fire} in Eastern New South Wales and the \emph{Cudlee Creek Fire} in Adelaide Hills, South Australia (Figure~\ref{fig:survey_location}), both of which started on or about December 18th, 2020. The Green Wattle Creek fire was extinguished by rainfall in February 2020 and is estimated to have destroyed 467,000 acres of land \citep{FiresWestSydney2019}. Australian firefighters put out the Cudlee Creek Fire, and damage was estimated at 57,000 acres \citep{AustralianWildfiresScorch}.

\subsection{Facebook population in Australia}

In Australia, it is estimated that there are 17.29 million people on Facebook \citep{fbAUStats} out of 25.4 million total people in Australia \citep{auCensus}; this provides a coverage of about 67 percent of the Australian population. Facebook skews slightly more females (53 percent; \citet{fbAUStats} ) compared to the Australian general population (51 percent; \citet{auCensus}). Facebook's age structure is heavily skewed towards those ages 18-50, with almost twice the percentage of the Australian population (see Table~\ref{tab:auDemographics}), with significantly fewer under 18-year-olds and people over 70 represented on Facebook as compared to the general population (see Table~\ref{tab:auDemographics})). This is similar to the United States, where Facebook usage also skews more females and younger than the general population \citep{pew}.

\begin{table}
\begin{longtable}{p{2cm}p{4.3cm}lll}
  \hline 
   &             & Australia & Australia Census \\
   & & FB Pop. Percentiles & Pop. Percentiles\\
     \hline
  Gender & &&\\
  & Female & 53.3 & 50.7\\
  &  Male   & 46.7 &  49.3\\
  Age  & &&\\
  &17$<$ &  03.8 &  22.0\\
 &18-19 &  04.6 &   02.0\\
 &  20-29 &   24.9 & 14.0  \\
 &  30-39 &   22.6 &  14.0\\
 &  40-49 &   16.5 &  13.0\\
 &  50-59 &   12.8 &  12.0\\
 &   60-69 &   09.4 &  10.0\\
 &   70+ &      05.5  &  11.0 \\
   \hline
\hline
\caption{Age and gender demographics for Australia Facebook \citep{fbAUStats}  and general population \citep{auCensus}.} 
\label{tab:auDemographics}
\end{longtable}
\end{table}

In the next several subsections, we discuss the Australian bushfires, the sampling frame, and the survey.

\section{Facebook survey of displaced populations due to Bushfires: Green Wattle Creek Fire in Eastern New South Wales and the Cudlee Creek Fire in Adelaide Hills, South Australia }

Using Facebook's internal survey infrastructure, we surveyed a stratified random sample of 24,486 Facebook users over 18. Each of these respondents answered various questions on demographics, displacement, smoke masks, and whether they were in the region at the time of the fires.  This was a large-scale pilot survey to demonstrate the efficacy of this approach, with 1,037,214 invitations and 96,414 starts. This provides a rough estimate of a response rate of 9.3\%.\footnote{Facebook survey infrastructure is an internal tool similar to Qualtrics or Survey Monkey and was administered at no cost to the researchers.} We stratified by Facebook users with location history turned on and in the region (similar to  \citet{maas2019facebook}) and a random sample of Facebook users predicted to be in the region (prediction is an internal product not available to researchers). Here, we are focusing on two bushfires that started on or about December 18, 2019 -- the \emph{Green Wattle Creek Fire} in Eastern New South Wales and the \emph{Cudlee Creek Fire} in Adelaide Hills, South Australia (Figure~\ref{fig:survey_location}). Of those who said they were in the area, 7,073 users said they were affected by the fires (Survey Appendix: Survey Questions and Logic: 5).  These users were then asked detailed questions on displacement outcomes. 

We limit our displacement-related analysis to these 7,073 users to guarantee we only consider the target population.  This is a form of rejection sampling and will maintain our sample properties \citep{gilks1992adaptive}. The survey was in the field (i.e., the survey was live and available to be taken by the requested survey respondents) for two weeks, from February 20, 2020, to March 5, 2020, approximately two months after the bushfires began.  This timing strikes a balance between decaying respondent recall and allowing the consequences of a displacement to play out more fully.

Generally speaking, this can be thought of as an account-based sampling frame for the Facebook population pre-disaster, where we continue to be able to survey people even if they are displaced or evacuated due to a disaster. This should be a satisfactory approximation of the general Australian population, given Facebook coverage of about 67\%. 

\subsection{Sampling Design}

\begin{figure}
\centering
  \includegraphics[width=\linewidth]{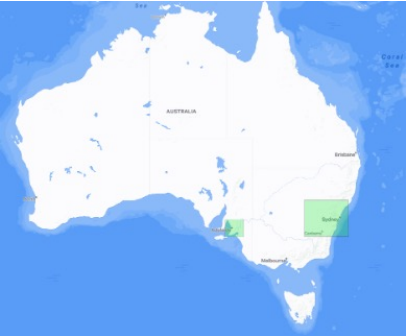}
\caption{Map of the bounding box survey locations: \emph{Green Wattle Creek Fire} in Eastern New South Wales and the \emph{Cudlee Creek Fire} in Adelaide Hills South Australia.}
\label{fig:survey_location}       
\end{figure}

The sampling design is built to capture a representative sample of Facebook users who were 18 or older and predicted to be in the disaster region of the \emph{Green Wattle Creek} and \emph{Cudlee Creek} fires (Figure~\ref{fig:survey_location}).  To obtain our final sample, we use a combination of internal targets and \emph{survey gatekeeper} questions. First, we include all Facebook users identified as possibly in the region by Location History records as reported in Facebook's Data for Good program \emph{Displacement Map} -- see \citet{maas2019facebook} for details. This subset of Facebook users has a smartphone and opt-in to allow Facebook access to the respondent's geolocation (and is generally about 15\% of the Facebook population \citep{maas2019facebook}). We augment this population with a random sample of users identified with an internal Facebook analysis as potentially located within the region of interest.\footnote{A subset of Facebook users opt into providing Location History data, and this detailed location information is used to determine if they are in the affected region to be included in the sample.  For Australian users without precise location information, we rely on a broader city prediction to include people from cities in the affected region, broadening the survey coverage beyond just Location History users.} This provides a stratified sample of respondents -- one major feature of this design is its ability to be compared with Facebook \emph{Displacement Maps} (which only include the geocoded users). To guarantee all users were in the natural hazard event region in the period of interest, we asked them specifically in the survey if they were in the Green Wattle Creek Fire in Eastern New South Wales and  Cudlee Creek Fire Adelaide Hills, South Australia, during the Bushfires that started on December 18th (Survey Appendix: Survey Questions and Logic: 2 and 3). 

\begin{figure}
\centering
  \includegraphics[width=\linewidth]{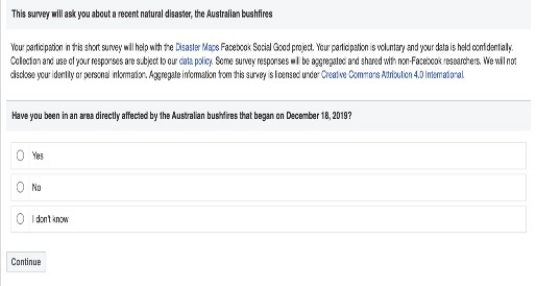}
\caption{Example survey question from the Facebook Australia Bushfire survey on the platform.}
\label{fig:survey_example}       
\end{figure}

Data are collected through a survey invitation shown to users at the top of their Facebook News Feed once eligibility for the survey has been established. We then employed basic rejection sampling to limit our final respondent set to include only the target population (i.e., anyone who answered they were not in the region during the disaster event and was not asked about displacement). This method is very general and provides a random sample of users in the two regions \citep{gilks1992adaptive}. We adjust for non-response by applying survey weights (see Section~\ref{weights}). All users were asked for informed consent to the survey, and all questions allowed the respondent not to respond (See Question 1).  No participants were compensated in any way for their participation. An example survey question can be seen in Figure~\ref{fig:survey_example}.

\subsection{Non-response and sample weights} 
\label{weights}

We employ calibration (raking) weights to allow for analysis representative of the over-18 Facebook population for the two regions under study. To do this, we weight on \emph{age, gender, engagement} (number of days active on Facebook in a month -- highly predictive of survey response), \emph{region}, and if the user has \emph{location history} turned on. This improves the representativeness of our results among core demographics and other key Facebook user categories. All statistics discussed in this article are re-weighted and employ the Horvitz-Thompson estimator for computing the weighted mean with standard errors computed using the delta method \citep{overton1995horvitz}. 

Like most surveys, we find that a large percentage of respondents, 47.9\%, preferred not to share income information. This is typical for surveys in developed countries, which neither require responses to any given question nor provide incentives for survey responses.  See \citep{riphahn2005item} for a good overview of item non-response on financial questions from the literature.  This paper suggests that there is a differential item of non-response and that refusal on questions of labor income is typical, but that it is somewhat mitigated for self-administered surveys.  However, we observed a much higher rate than this paper, which we attribute to the format of data collection and study participation.  We expect that Facebook will fall somewhere between interviewer-administered surveys we could have run and self-administered surveys run by other institutions, as Facebook is a large public platform.  However, we do not have a strong prior on the relative magnitude of that effect. Our survey weighting process further ameliorates this using basic demographics and usage statistics. A full comparison of the post-weighted data follows.

\subsection{Demographics of the Facebook survey}

In Table~\ref{demographics}, we compare the Facebook survey gender and age demographics with that of the Facebook platform in percentiles and Australian Census estimates. As expected, the Facebook survey marginals and the Facebook population marginals are highly comparable. Major differences in the Facebook sample and the Australian population, in general, include approximately 4\% more 18-19-year-olds than expected, 1-2\% more 20-29 and 30-39-year-olds;  comparable 40-49-year-olds, but 3\% less 60-69-year-olds, and almost 10\% less 70+-year-olds. Thus, the survey skewed slightly younger and a bit more female (2\%) than the general Australian population.  We adjust these statistics using the weights discussed in Section~\ref{weights}. 

\begin{table}
\begin{longtable}{p{2cm}p{4.3cm}llll}
  \hline 
  & & & &FB  &Australia Census\\
Demographic & Response & N & Percent in Sample & Pop. Percentiles & Pop. Percentiles \\ 
  \hline
Reported Gender &  &     & &  \\ 
   & Female & 42230 & 54.2 (53.6, 54.7) & 53.30 & 50.7 \\ 
   & Male & 35717 & 45.8 (45.3, 46.4) & 46.70 & 49.3 \\ 
  Reported Age &  &     &  \\ 
   & 18-19 &  6171 &  7.7 ( 7.4,  8.1) & 4.63 & 3.00 \\
   & 20-29 & 17984 & 22.6 (22.2, 23.0) & 24.87 & 19.00 \\
   & 30-39 & 16200 & 20.3 (19.9, 20.8) &  22.55 & 19.00 \\
   & 40-49 & 12971 & 16.3 (15.9, 16.7) &  16.50 & 17.00 \\
   & 50-59 & 10692 & 13.4 (13.0, 13.8) & 12.75 & 16.00\\
   & 60-69 &  7785 &  9.8 ( 9.4, 10.1) & 9.42 & 13.00 \\
   & 70+ &  4643 &  5.8 ( 5.5,  6.1) &  5.48 & 14.00 \\
   & Prefer Not to Say &  3200 &  4.0 ( 3.8,  4.3) & &\\ 
 \hline
\hline
\caption{Comparison of FB Survey Age and Gender to Facebook Australia demographics  \citep{fbAUStats}  and Australia Demographics \citep{ParticipationJobSearch2020} estimates for January, 2020. For these statistics, we adjust using the weights discussed in Section~\ref{weights}.} 
\label{demographics}
\end{longtable}
\end{table}

Next, we compared the Facebook survey to a core set of comparable statistics in the Australian population: households of size one, employment status, and education, see Table~\ref{hemployeduc}. We use the household size of one estimate from the Australia census \citep{auCensus} where we see the survey is about 3\% points smaller than the Australian household size of one estimate. We use the Australia Bureau of Statistics estimate of education \citep{EducationWorkAustralia2019} to compute education statistics for Australia for comparison with our survey. We find that the Facebook survey is about 2.8\% lower in junior high graduates than expected; high school and community college is 51\% on the Facebook survey versus 47\% in the general population (so about 4\% higher than expected). The Facebook survey is 2\% lower in university graduates (17\% versus 19\%), but 4\% points higher in greater than university degree (14\% versus 10\%). Overall, the Facebook population is a little more educated than the wider Australian population. 

Finally, we compare employment to the Australian BLS \citep{ParticipationJobSearch2020} and the count of students  \citep{StudentStatistics}. We find that our sample is more likely to be employed (57\% compared to 51\%), slightly more likely to be a student (7.7\% compared to 5.5\%), less likely to be retired (11.5\% compared to 15\%), and more likely to report being unemployed (11.2\% compared to 8.2\%). We adjust descriptive statistics using the weights discussed in Section~\ref{weights}. 

\begin{table}
\begin{longtable}{p{2cm}p{4.3cm}lll}
  \hline 
  & & & & Australia Census\\
Demographic & Response & N & Percent in Sample & Pop. Percentiles \\ 
  \hline
  Household Size &  &     &  & \\ 
   & 1 &  8964 & 12.3 (11.8, 12.7) & 15.8 \\ 
Education &  &     &  \\ 
   & Elementary &   663 &  2.0 ( 1.7,  2.4) &  -- \\ 
   & Junior High &  1291 &  4.0 ( 3.6,  4.4) & 6.8\\ 
   & High School &  9417 & 29.1 (28.3, 29.9) & 20.0\\ 
   & Community College &  6996 & 21.6 (20.9, 22.3) & 27.0  \\ 
   & University &  5554 & 17.2 (16.5, 17.8) &  19.00\\ 
   & Graduate School &  4750 & 14.7 (14.1, 15.3) & 10.00  \\ 
   & Prefer Not to Say &  3711 & 11.5 (10.9, 12.1) &  --\\ 
  Employment Type &  &     &  \\ 
   & Employed & 39525& 57.0 (57.1, 58.0) & 51.0\\
   & Student &  5329 &  7.7 ( 7.4,  8.1) & 5.5\\ 
   & Retired &  7898 & 11.5 (11.0, 11.9) & 15.0\\ 
   & Not Working &  7689 & 11.2 (10.7, 11.6) & 8.2 (unemployed) \\ 
   & Prefer Not to Say &  8484 & 12.3 (11.9, 12.7) & \\ 
   \hline
\hline
\caption{Comparison of Household size of one \citep{auCensus}, Education \citep{EducationWorkAustralia2019} and employment \citep{ParticipationJobSearch2020} with student number coming from \citet{StudentStatistics}.}
\label{hemployeduc}
\end{longtable}
\end{table}

Finally, we have a table of all descriptive statistics and demographics from the survey in Table~\ref{survey}.

\begin{table}
\begin{longtable}{p{2cm}p{4.3cm}lll}
  \hline 
Demographic & Response & N & Percent in Sample \\ 
  \hline
Reported Gender &  &     & &  \\ 
   & Female & 42230 & 54.2 (53.6, 54.7) &  \\ 
   & Male & 35717 & 45.8 (45.3, 46.4) & \\ 
  Reported Age &  &     &  \\ 
   & $<$20 &  6171 &  7.7 ( 7.4,  8.1) & \\ 
   & 20-29 & 17984 & 22.6 (22.2, 23.0) & \\ 
   & 30-39 & 16200 & 20.3 (19.9, 20.8) & \\ 
   & 40-49 & 12971 & 16.3 (15.9, 16.7) & \\ 
   & 50-59 & 10692 & 13.4 (13.0, 13.8) & \\ 
   & 60-69 &  7785 &  9.8 ( 9.4, 10.1) & \\ 
   & 70+ &  4643 &  5.8 ( 5.5,  6.1) &  \\ 
   & Prefer Not to Say &  3200 &  4.0 ( 3.8,  4.3) & \\ 
  Education &  &     &  \\ 
   & Elementary &   663 &  2.0 ( 1.7,  2.4) &  \\ 
   & Junior High &  1291 &  4.0 ( 3.6,  4.4) & \\ 
   & High School &  9417 & 29.1 (28.3, 29.9) & \\ 
   & Community College &  6996 & 21.6 (20.9, 22.3) &  \\ 
   & University &  5554 & 17.2 (16.5, 17.8) &  \\ 
   & Graduate School &  4750 & 14.7 (14.1, 15.3) &  \\ 
   & Prefer Not to Say &  3711 & 11.5 (10.9, 12.1) &  \\ 
  Employment Type &  &     &  \\ 
   & Managing a Business &  5506 &  8.0 ( 7.7,  8.3) & \\ 
   & Employed by Business & 25350 & 36.8 (36.2, 37.3) & \\ 
   & Employed, not by Business &  2686 &  3.9 ( 3.7,  4.1) & \\ 
   & Government Work &  5983 &  8.7 ( 8.4,  9.0) & \\ 
   & Student &  5329 &  7.7 ( 7.4,  8.1) & \\ 
   & Retired &  7898 & 11.5 (11.0, 11.9) & \\ 
   & Not Working &  7689 & 11.2 (10.7, 11.6) & \\ 
   & Prefer Not to Say &  8484 & 12.3 (11.9, 12.7) & \\ 
  Income &  &     &  &  \\ 
   & $<$ \$3,200 &  5892 & 18.9 (18.1, 19.7) & \\ 
   & \$3,200-\$5,800 &  4263 & 13.7 (13.1, 14.3) & \\ 
   & \$5,800-\$9,100 &  2877 &  9.2 ( 8.8,  9.7) & \\ 
   & \$9,100-\$14,000 &  1547 &  5.0 ( 4.6,  5.3) & \\ 
   & $>$ \$14,000 &  1624 &  5.2 ( 4.8,  5.6) & \\ 
   & Prefer Not to Say & 14918 & 47.9 (47.0, 48.8) & \\ 
  Household Head &  &     &  & \\ 
   & Yes & 24727 & 34.9 (34.3, 35.4) & \\ 
   & No & 31150 & 43.9 (43.3, 44.5) & \\ 
   & Prefer not to say & 15062 & 21.2 (20.7, 21.7) & \\ 
  Household Size &  &     &  & \\ 
   & 1 &  8964 & 12.3 (11.8, 12.7) & \\ 
   & 2 & 20319 & 27.8 (27.3, 28.3) & \\ 
   & 3 & 14392 & 19.7 (19.2, 20.1) & \\ 
   & 4 & 14910 & 20.4 (19.9, 20.8) & \\ 
   & 5 &  8197 & 11.2 (10.8, 11.6) & \\ 
   & 6 or more &  6360 &  8.7 ( 8.4,  9.0) & \\ 
  Household Child Count &  &     &  & \\ 
   & 0 & 42914 & 58.7 (58.1, 59.2) & \\ 
   & 1 & 12010 & 16.4 (16.0, 16.8) & \\ 
   & 2 & 10701 & 14.6 (14.2, 15.0) & \\ 
   & 3 &  4424 &  6.0 ( 5.8,  6.3) & \\ 
   & 4 &  1653 &  2.3 ( 2.1,  2.4) & \\ 
   & 5 or more &  1453 &  2.0 ( 1.8,  2.2) & \\ 
  Household Partner &  &     &  \\ 
   & Yes & 37940 & 53.2 (52.6, 53.8) & \\ 
   & No & 26162 & 36.7 (36.1, 37.3) & \\ 
   & Prefer not to say &  7163 & 10.1 ( 9.7, 10.4) & \\ 
   \hline
\hline
\caption{Descriptive statistics of the FB Survey for the Australian Bushfires.} 
\label{survey}
\end{longtable}
\end{table}

\subsection{Estimating the Total Number of Displaced People for Adelaide  Hills, South  Australia}

 
Given that we are working from a probability sample, we can apply design-based estimation such as the Horvitz-Thompson (HT) estimator \citep{lumley2004analysis} to \emph{estimate} the total number of people displaced from the disaster-affected area. The HT estimator for total is $T_{HT} = N\cdot \hat{p}_{displaced}$. To estimate the total, we need to estimate $\hat{N}$ by using the Australia Census estimate for the Adelaide  Hills, South  Australia area, which is $\hat{N} = 71,892$ \citep{auCensus}. If we apply the weighted estimate of the displaced population ($\hat{p}_{displaced} = 0.25$) from our survey, we can estimate the total number of individuals displaced for more than one night two months post-Cudlee Creek Fire. This provides us the $\hat{T}_{displaced} = 17,973$ with a 95\% Confidence Interval of 17,741 to 18,205 individuals.

We can compare this estimate to that of the International Displaced Monitoring Centre \cite{20192020AUSTRALIANBUSHFIRES} estimates of about 17,000 people displaced from January to February in the Adelaide Hills, South Australia; this estimate is from Red Cross Data. Our estimate is not wildly different from the IDMC estimate, about $\sim 741-1,205$ more people. This seems largely in line with expectations, given the discussion by IDMC on their data source being a likely underestimation.

\section{Analysis}

 We limit our analysis to the affected areas covered by our survey (the \emph{Green Wattle Creek Fire} in Eastern New South Wales and the \emph{Cudlee Creek Fire} in Adelaide Hills, South Australia (Figure~\ref{fig:survey_location}). To guarantee we only consider the target population, we limit our displacement-related analysis to these 7,073 users who met the full criteria of being in the sampling frame and passing the \emph{gatekeeper} questions (i.e., verified they were in the region during the disaster event).
 
 Our analysis proceeds topic-by-topic, with commentary on potential implications for policy response to disaster displacement.

\subsubsection{Who evacuates?}
\label{sec:5}

We found that among the 7,073 respondents in the affected areas, 25.5\%, reported being displaced more than one night. Using a logistic regression model (Table~\ref{Total_weights_q11_one_night_regression_no_interaction_with_income}), we find that odds for being displaced or evacuated more than one night are lower among those aged 20 to 29 and higher among older respondents (50-59, 60-69 and 70+).  We observe that those with higher incomes were less likely to be displaced than those with lower incomes. Finally, being head of a household is associated with not being displaced.

\begin{table}
\begin{longtable}{llrlll}
  \hline
Response & Covariate & Odds Ratio & 95\% CI & P Value & Significance \\ 
  \hline
No & Gender: Male & 1.00 &  &  &  \\ 
  No & Gender: Female & 0.92 & (0.80, 1.1) & 0.266 &   \\ 
  No & Gender: Prefer Not to Say & 1.00 & (1.00, 1.0) & NaN &  \\ 
  No & Age: 20-29 & 0.82 & (0.68, 1.0) & 0.043 & * \\ 
  No & Age: 30-39 & 1.00 &  &  &  \\ 
  No & Age: 40-49 & 0.95 & (0.77, 1.2) & 0.601 &   \\ 
  No & Age: 50-59 & 1.52 & (1.19, 2.0) & 0.001 & *** \\ 
  No & Age: 60-69 & 1.30 & (0.98, 1.7) & 0.069 & . \\ 
  No & Age: 70+ & 1.07 & (0.73, 1.6) & 0.737 &   \\ 
  No & Age: Prefer Not to Say & 0.76 & (0.50, 1.2) & 0.216 &   \\ 
  No & Income: $<$ \$3,200 & 1.00 &  &  &  \\ 
  No & Income: \$3,200-\$5,800 & 1.43 & (1.04, 2.0) & 0.026 & * \\ 
  No & Income: \$5,800-\$9,100 & 1.87 & (1.30, 2.7) & 0.001 & *** \\ 
  No & Income: \$9,100-\$14,000 & 1.86 & (1.13, 3.1) & 0.015 & * \\ 
  No & Income: $>$ \$14,000 & 2.16 & (1.30, 3.6) & 0.003 & ** \\ 
  No & Income: Prefer Not to Say & 1.90 & (1.46, 2.5) & 0.000 & *** \\ 
  No & Household\_head: Yes & 1.00 &  &  &  \\ 
  No & Household\_head: No & 1.32 & (1.12, 1.5) & 0.001 & *** \\ 
   \hline
\hline
\caption{Displaced more than one night, compared to ``yes"; relative to men, aged 30-39, income less than \$3,200 head of household based on a model with main effects for age, gender, income, and head of household with no interactions} 
\label{Total_weights_q11_one_night_regression_no_interaction_with_income}
\end{longtable}
\end{table}

\subsubsection{Who makes the decision to evacuate?}
\label{sec:6}

\begin{figure}
  \includegraphics[width=\linewidth]{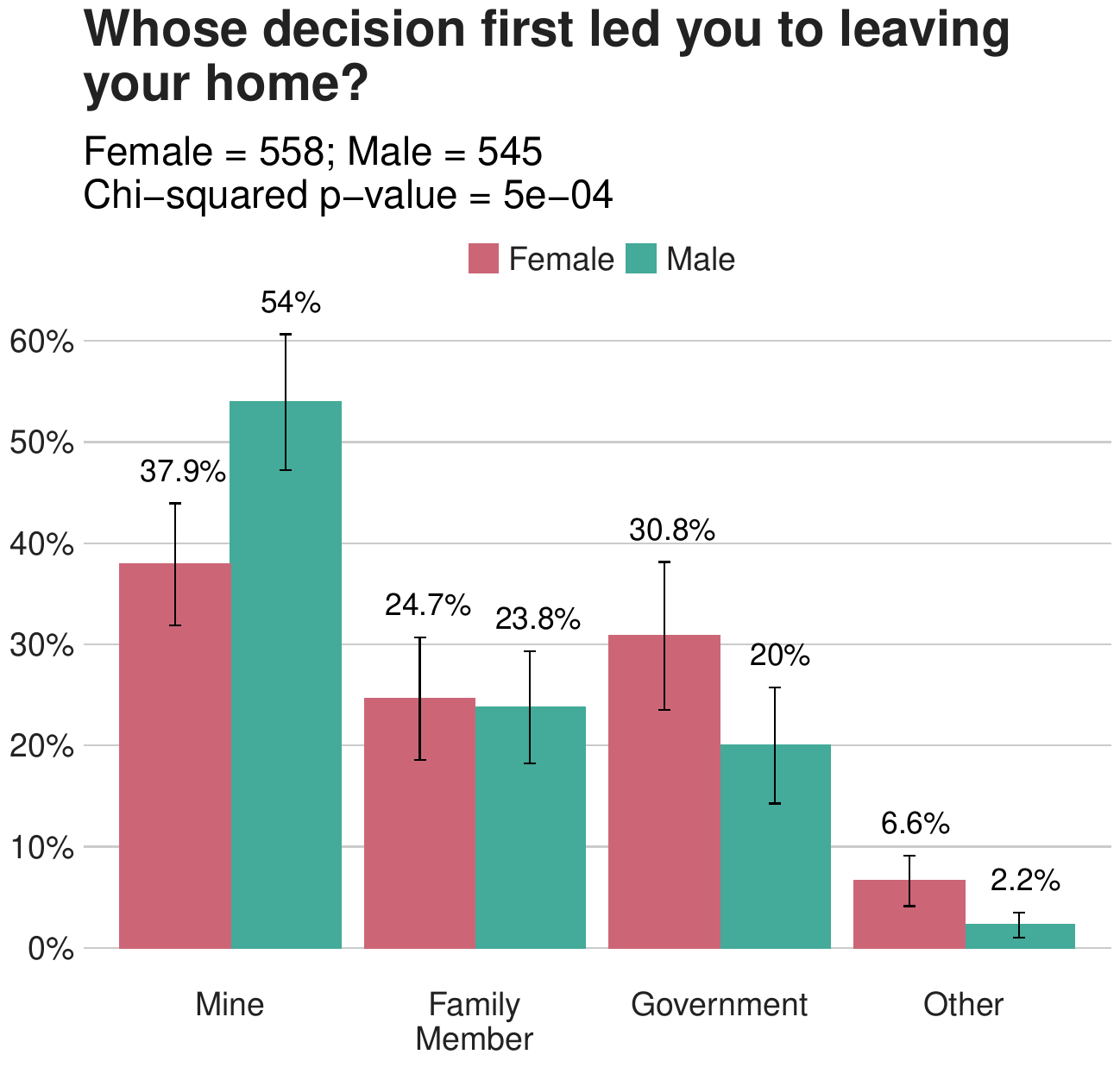}
\caption{Men are significantly more likely to say that the decision to evacuate was their own, while women are more likely to attribute it to the government.}
\label{fig:2}       
\end{figure}

Of those respondents who were asked, “Whose decision first led to you leaving your home?”  (Survey Appendix: Survey Questions and Logic: 19), the plurality responded that it was their own decision (47.7\%), while 24.3\% and 23.6\% attributed the decision to a family member or the government, respectively.  Men were significantly more likely to say it was their own decision, while women were more likely to say it was a government decision based on a $\chi^2$ test (Figure \ref{fig:2}). We caution that these attribution decisions may be subject to some self-reporting bias.  

This result holds when controlling for gender, age, and education and whether the person was the head of their household (Table~\ref{Total_weights_q29_evacuation_decision_regression_no_interaction}). We find an odds ratio of 2.3 for women attributing the decision to their government as compared to attributing it to themselves (statistically significant at the $\alpha=0.05$-level). In these models, we find that people who were not head of household had twice the odds of attributing the evacuation decision to a family member. 

\begin{table}
\begin{longtable}{llrlll}
  \hline
Response & Covariate & Odds Ratio & 95\% CI & P Value & Significance \\ 
  \hline
Family Member & Gender: Male & 1.00 &  &  &  \\ 
  Family Member & Gender: Female & 1.28 & (0.91, 1.8) & 0.149 &   \\ 
  Family Member & Age: 20-29 & 2.28 & (1.48, 3.5) & 0.000 & *** \\ 
  Family Member & Age: 30-39 & 1.00 &  &  &  \\ 
  Family Member & Age: 40-49 & 1.57 & (0.95, 2.6) & 0.077 & . \\ 
  Family Member & Age: 50-59 & 0.71 & (0.36, 1.4) & 0.314 &   \\ 
  Family Member & Age: 60-69 & 0.33 & (0.11, 1.0) & 0.052 & . \\ 
  Family Member & Age: 70+ & 1.74 & (0.67, 4.5) & 0.252 &   \\ 
  Family Member & Education: Elementary & 2.58 & (0.48, 13.7) & 0.267 &   \\ 
  Family Member & Education: Junior High & 2.25 & (0.65, 7.8) & 0.201 &   \\ 
  Family Member & Education: High School & 1.58 & (0.78, 3.2) & 0.204 &   \\ 
  Family Member & Education: Community College & 0.79 & (0.35, 1.8) & 0.567 &   \\ 
  Family Member & Education: University & 1.00 &  &  &  \\ 
  Family Member & Education: Graduate School & 2.01 & (0.93, 4.4) & 0.077 & . \\ 
  Family Member & Household\_head: Yes & 1.00 &  &  &  \\ 
  Family Member & Household\_head: No & 2.03 & (1.40, 2.9) & 0.000 & *** \\ 
  Government & Gender: Male & 1.00 &  &  &  \\ 
  Government & Gender: Female & 2.34 & (1.68, 3.3) & 0.000 & *** \\ 
  Government & Age: 20-29 & 1.77 & (1.13, 2.8) & 0.013 & * \\ 
  Government & Age: 30-39 & 1.00 &  &  &  \\ 
  Government & Age: 40-49 & 1.92 & (1.19, 3.1) & 0.007 & ** \\ 
  Government & Age: 50-59 & 0.94 & (0.51, 1.7) & 0.832 &   \\ 
  Government & Age: 60-69 & 1.87 & (1.01, 3.5) & 0.048 & * \\ 
  Government & Age: 70+ & 3.21 & (1.41, 7.3) & 0.005 & ** \\ 
  Government & Education: Elementary & 2.23 & (0.49, 10.2) & 0.301 &   \\ 
  Government & Education: Junior High & 1.51 & (0.41, 5.5) & 0.536 &   \\ 
  Government & Education: High School & 0.55 & (0.26, 1.2) & 0.115 &   \\ 
  Government & Education: Community College & 1.57 & (0.81, 3.0) & 0.178 &   \\ 
  Government & Education: University & 1.00 &  &  &  \\ 
  Government & Education: Graduate School & 0.78 & (0.34, 1.8) & 0.545 &   \\ 
  Government & Household\_head: Yes & 1.00 &  &  &  \\ 
  Government & Household\_head: No & 0.65 & (0.45, 0.9) & 0.022 & * \\ 
   \hline
\hline
\caption{Evacuation decision made by someone else, compared to ``my decision"; relative to university-educated men, aged 30-39, head of household based on the model with main effects for age, gender, education, and head of household with no interactions.} 
\label{Total_weights_q29_evacuation_decision_regression_no_interaction}
\end{longtable}
\end{table}

\subsubsection{Where do people go when they evacuate?}
\label{sec:7}

\begin{figure}
  \includegraphics[width=\linewidth]{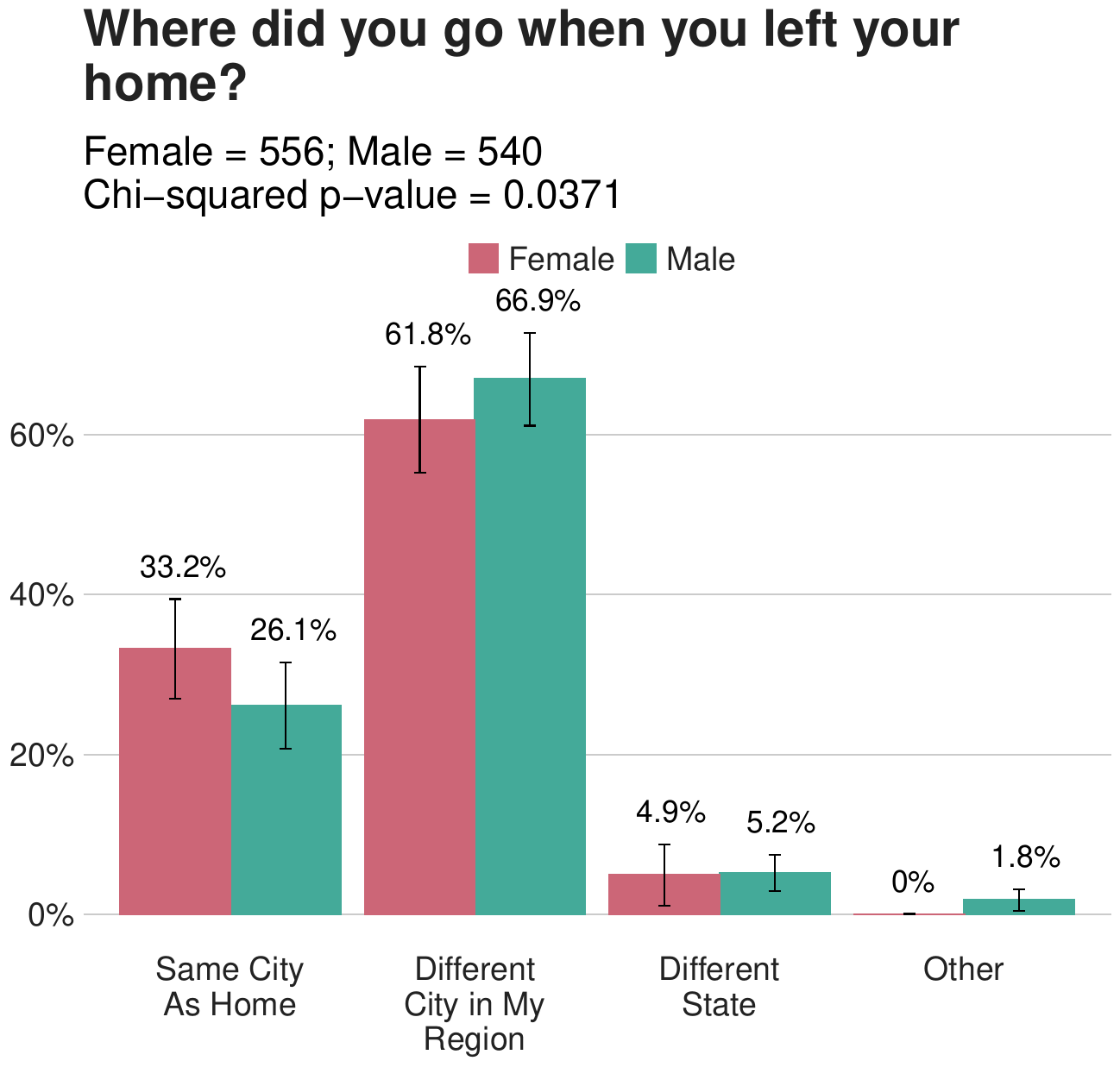}
\caption{Women are more likely to evacuate within the same city as their home.  This corresponds with trends in the Displacement Map data across many disasters and geographies.}
\label{fig:3}       
\end{figure}

Of people who were evacuated or displaced more than one night, the majority (63.4\%) evacuated to a different city in their region. At the same time, a much smaller fraction went somewhere else in their home city (30\%).  A small percentage went to a different state in Australia (5.4\%) (see Figure~\ref{fig:3} for more detail). Women were significantly more likely to report that they evacuated to another location in their home city based on $\chi^2$-test (Figure \ref{fig:3}).  These findings suggest that proper policy responses should account for a significant increase in need in locations near, but not in, affected disaster areas.  Furthermore, the finding of differences by gender implies at least some differences in destination conditional on household separation or household structure. We can infer that policy responses may need to be tailored to the populations expected to relocate given distances. 

\subsubsection{Under what circumstances do households separate during displacement?}
\label{sec:8}

\begin{figure}
  \includegraphics[width=\linewidth]{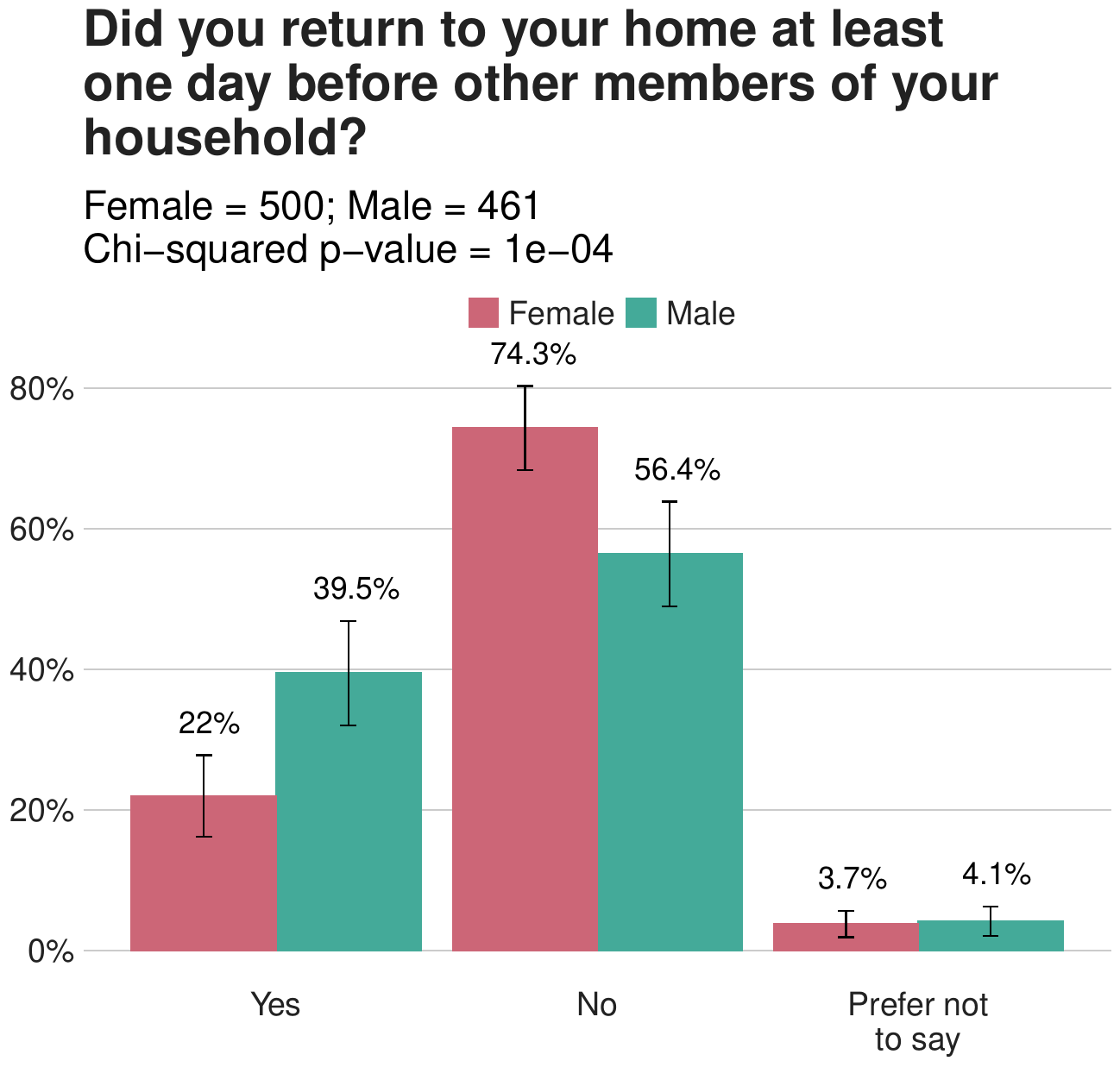}
\caption{Men are significantly more likely to return home ahead of the other household members.}
\label{fig:4}       
\end{figure}

To understand the extent to which households were separated during displacement, we asked two questions: (1) ``When you were displaced for more than three nights did anyone in your household stay behind?" (Survey Appendix: Survey Questions and Logic: 10) and ``Did you return home at least one day before other members of your household?" (Survey Appendix: Survey Questions and Logic: 22). We find that a substantial portion of households do separate either at the beginning (28\%) or end (31.2\%) of their displacement (see Figure~\ref{fig:4}).  Interestingly, the responses to these two questions are uncorrelated (at the $\alpha=0.05$ level based on $\chi^2$ test), which is to say that households that separated during departure are no more or less likely to return separately than those that did not. We did not observe gender differences for households separating upon departure. However, we found that men return home sooner than their household more frequently than women (Figure \ref{fig:4}).  

In a regression model adjusted for age, gender, education, head of household status, and household structure: men (OR = 2.1), people aged 50-59 (OR = 2), heads of household (OR = 2.4) and people with any children (OR = 1.6) are the groups more likely to return home ahead of their households (Table \ref{Total_weights_q26_household_split_return_no_ref_regression_no_interaction_with_child_count_reformulated}).  People without a university degree are significantly less likely to return home ahead of others in their household.

\begin{table}
\begin{longtable}{llrlll}
  \hline
Response & Covariate & Odds Ratio & 95\% CI & P Value & Significance \\ 
  \hline
Yes & Gender: Male & 2.07 & (1.51, 2.9) & 0.000 & *** \\ 
  Yes & Gender: Female & 1.00 &  &  &  \\ 
  Yes & Gender: Prefer Not to Say & 1.00 & (NaN, NaN) & NaN &  \\ 
  Yes & Age: 20-29 & 1.19 & (0.75, 1.9) & 0.462 &   \\ 
  Yes & Age: 30-39 & 1.00 &  &  &  \\ 
  Yes & Age: 40-49 & 1.36 & (0.87, 2.1) & 0.182 &   \\ 
  Yes & Age: 50-59 & 2.10 & (1.19, 3.7) & 0.011 & * \\ 
  Yes & Age: 60-69 & 0.62 & (0.28, 1.4) & 0.246 &   \\ 
  Yes & Age: 70+ & 2.50 & (1.07, 5.9) & 0.035 & * \\ 
  Yes & Age: Prefer Not to Say & 1.80 & (0.73, 4.4) & 0.202 &   \\ 
  Yes & Education: Elementary & 0.48 & (0.04, 5.7) & 0.564 &   \\ 
  Yes & Education: Junior High & 0.48 & (0.15, 1.5) & 0.216 &   \\ 
  Yes & Education: High School & 0.39 & (0.21, 0.7) & 0.004 & ** \\ 
  Yes & Education: Community College & 0.34 & (0.18, 0.7) & 0.001 & ** \\ 
  Yes & Education: University & 1.00 &  &  &  \\ 
  Yes & Education: Graduate School & 0.09 & (0.03, 0.2) & 0.000 & *** \\ 
  Yes & Education: Prefer Not to Say & 0.51 & (0.20, 1.3) & 0.149 &   \\ 
  Yes & Household\_head: Yes & 2.38 & (1.66, 3.4) & 0.000 & *** \\ 
  Yes & Household\_head: No & 1.00 &  &  &  \\ 
  Yes & Any\_children: FALSE & 1.00 &  &  &  \\ 
  Yes & Any\_children: TRUE & 1.65 & (1.17, 2.3) & 0.004 & ** \\ 
   \hline
\hline
\caption{Return home sooner, compared to ``no"; relative to university-educated women, aged 30-39, not head of household, with no children based on the model with main effects for age, gender, education, and head of household with no interactions. } 
\label{Total_weights_q26_household_split_return_no_ref_regression_no_interaction_with_child_count_reformulated}
\end{longtable}
\end{table}

\subsubsection{Whose work is disrupted?}
\label{sec:9}

We also assess the potential economic impact of displacement through employment disruption.  Among displaced people asked: “Did leaving your home prevent you from working as much as you normally do?” 54.7\% of people said “yes.”  These responses did not significantly differ by gender ($\alpha=0.05$-level). To better understand patterns of work disruption associated with displacement, we employed a logistic regression model controlling for gender, age, income, household head status, and displacement length (1-3 nights and more than 3 nights); see Table~\ref{Total_weights_q21_prevent_working_regression_no_interaction_with_income_displacement_reformulated}. The major finding is that being displaced more than 3 nights increases ones odds of being prevented from working (OR = 1.8). This is statistically significant at the $\alpha=0.05$-level. We find significantly lower impacts among older respondents, likely due to a higher incidence of being retired. These are relatively large impacts, and further work could help quantify better the economic costs associated with natural disasters.

\begin{table}
\begin{longtable}{llrlll}
  \hline
Response & Covariate & Odds Ratio & 95\% CI & P Value & Significance \\ 
  \hline
Yes & Gender: Male & 1.00 &  &  &  \\ 
  Yes & Gender: Female & 0.91 & (0.68, 1.2) & 0.542 &   \\ 
  Yes & Gender: Prefer Not to Say & 1.00 & (1.00, 1.0) & NaN &  \\ 
  Yes & Age: 20-29 & 1.42 & (0.96, 2.1) & 0.082 & . \\ 
  Yes & Age: 30-39 & 1.00 &  &  &  \\ 
  Yes & Age: 40-49 & 1.05 & (0.69, 1.6) & 0.826 &   \\ 
  Yes & Age: 50-59 & 1.07 & (0.64, 1.8) & 0.802 &   \\ 
  Yes & Age: 60-69 & 0.27 & (0.14, 0.5) & 0.000 & *** \\ 
  Yes & Age: 70+ & 0.58 & (0.25, 1.3) & 0.184 &   \\ 
  Yes & Age: Prefer Not to Say & 1.55 & (0.52, 4.6) & 0.437 &   \\ 
  Yes & Income: $<$ \$3,200 & 1.00 &  &  &  \\ 
  Yes & Income: \$3,200-\$5,800 & 1.66 & (0.86, 3.2) & 0.131 &   \\ 
  Yes & Income: \$5,800-\$9,100 & 0.68 & (0.34, 1.4) & 0.276 &   \\ 
  Yes & Income: \$9,100-\$14,000 & 2.20 & (0.69, 7.0) & 0.183 &   \\ 
  Yes & Income: $>$ \$14,000 & 0.88 & (0.32, 2.4) & 0.798 &   \\ 
  Yes & Income: Prefer Not to Say & 0.47 & (0.28, 0.8) & 0.006 & ** \\ 
  Yes & Household\_head: Yes & 1.00 &  &  &  \\ 
  Yes & Household\_head: No & 1.21 & (0.88, 1.7) & 0.242 &   \\ 
  Yes & Displaced\_status: Displaced 1-3 nights & 1.00 &  &  &  \\ 
  Yes & Displaced\_status: Displaced more than 3 nights & 1.79 & (1.35, 2.4) & 0.000 & *** \\ 
   \hline
\hline
\caption{Prevent working, compared to ``no"; relative to men, aged 30-39, income less than \$3,200 head of household, not displaced based on the model with main effects for age, gender, income, head of household, and duration displaced with no interactions. Total sample 983.} 
\label{Total_weights_q21_prevent_working_regression_no_interaction_with_income_displacement_reformulated}
\end{longtable}
\end{table}

\subsubsection{For how long are people displaced?}
\label{sec:10}

We analyze the duration of displacement spells by asking respondents how long they were displaced after their return. 

We observe a strong, significant effect of family size, particularly having children, on the probability of being displaced for more than one week.   This effect is present even after controlling for education, gender, and household head status of respondents.  Those with 3 children were significantly more likely to be displaced more than one week (1.8 times as likely), and this effect increases with increasing household size (up to 4.4 times as likely for those reporting 5 or more children). See Table~\ref{tab:regmorethanweek}.


\begin{table}
\begin{longtable}{llrlll}
  \hline
Response & Covariate & Odds Ratio & 95\% CI & P Value & Significance \\ 
  \hline
Yes & Gender: Male & 1.00 &  &  &  \\ 
  Yes & Gender: Female & 1.08 & (0.77, 1.52) & 0.66 &   \\ 
  Yes & Gender: Prefer Not to Say & 1.00 & (1.00, 1.00) & NaN &  \\ 
  Yes & Age: 20-29 & 1.36 & (0.85, 2.18) & 0.20 &  \\ 
  Yes & Age: 30-39 & 1.00 &  &  &  \\ 
  Yes & Age: 40-49 & 1.24 & (0.73, 2.09) & 0.42 &   \\ 
  Yes & Age: 50-59 & 0.68 & (0.36, 1.30) & 0.24 &   \\ 
  Yes & Age: 60-69 & 1.21 & (0.56, 2.63) & 0.63 &  \\ 
  Yes & Age: 70+ & 1.52 & (0.51, 4.49) & 0.45 &   \\ 
  Yes & Age: Prefer Not to Say & 2.20 & (0.87, 5.56) & 0.10 & .   \\ 
  Yes & Education: Junior High & 0.76 & (0.22, 2.66) & 0.67 &   \\ 
  Yes & Education: High School & 1.60 & (0.82, 3.14) & 0.17 &   \\ 
  Yes & Education: Community College & 0.91 & (0.43, 1.94) & 0.81 &   \\ 
  Yes & Education: University & 1.00 &  &  &   \\
  Yes & Education: Graduate School & 1.71 & (0.77, 3.80) & 0.19 &   \\ 
  Yes & Education: Prefer Not to Say & 0.47 & (0.25, 1.62) & 0.35 & \\ 
  Yes & Household\_head: Yes & 1.00 &  &  &  \\ 
  Yes & Household\_head: No & 0.76 & (0.52, 1.10) & 0.15 &   \\ 
  Yes & Household\_child\_count: 0 & 1.00 &  &  &  \\ 
  Yes & Household\_child\_count: 1 & 0.85 & (0.54, 1.33)  & 0.47 &  \\ 
  Yes & Household\_child\_count: 2 & 1.58 & (1.00, 2.49) & 0.05 & *  \\ 
  Yes & Household\_child\_count: 3 & 1.84 & (1.05, 3.22) & 0.03 & * \\ 
  Yes & Household\_child\_count: 4 & 2.85 & (1.11, 7.34) & 0.03 & * \\ 
  Yes & Household\_child\_count: 5 or more & 4.41 & (1.75, 11.13) & 0.00  & *** \\ 
   \hline
\hline
\caption{Displaced more than 1 week, compared to displaced less than 1 week; relative to men, aged 30-39, university education, head of household, and no children with main effects for age, gender, income, head of household, and household size. Total sample size 561.} 
\label{tab:regmorethanweek}
\end{longtable}
\end{table}

\section{Discussion}

This survey, combined with geolocated displacement data, provides a timely and detailed resource on the effects of natural disasters on local populations, researchers, and policymakers. Policy responses to natural disasters benefit greatly from understanding the demographics and decisions people make on whether to evacuate and whether to return or not, and surveys such as this Facebook survey provide key insight into these actions. Social support can be better designed to account for likely household displacement and separation patterns.  Material responses can better adapt to the duration and volume of displacement.  Typical modes and timings of displacement events can inform transportation and logistical decisions.  Our findings also explore demographic differences in experiences of displacement, which merit further exploration. In particular, further research can explore whether the gender differences we find are related to gender norms surrounding household roles, differences in household composition and employment situation, or status by gender (all of which have been shown to be real phenomena in the Australian context).  The results are consistent with other work in the area, such as \citet{sastry2014location}.

Finally, this study provides a proof of concept for learning about the effects of fire on displacement, offering an investigation into heterogeneity in responses by demographic characteristics.  We find significant differences by gender and income on various outcomes, including evacuation timing, agency in evacuation decisions, and household separation.  This work reinforces the idea that disasters have highly heterogeneous effects on the affected population.

\subsection{Data limitations and the Facebook population}

These results, like all survey data, take at face value respondents' self-reporting as honest and are subject to potential measurement error. For example, respondents may attribute the decision to evacuate to their government or their own personal considerations should the government suggest, but not require, evacuation in a given area. These biases are, however, known to the survey literature \cite[e.g.][]{sastry2009tracing}, and the survey is designed to minimize them to the extent possible.

Furthermore, it is possible that the sampling frame used (Facebook users) is not totally representative of the general population. This is a general issue in survey science that results in some expected levels of error in results \footnote{One example comes from typical standards for most political polling, which have a margin of error between 3 and 4\% range\citep{jackman2005pooling}}. We compare our sample to government statistics and find that only areas of bias are excessive representation among those with graduate education (0.7 points over this threshold), employed persons (2 percentage points over this threshold), 18-19-year-olds (0.7 percentage points over this threshold), and an undersample of those age 70 and over  (4.6 percentage points under this threshold). These are well within the ranges normally considered correctable through survey weighting.

\subsubsection{Limitations on displacement timing}

Our survey was conducted two months after the beginning of the 2019 Bushfires and is thus limited in its ability to measure longer-term displacement.  However, we can assess many characteristics of displacement caused by this event, which can be seen in Figure~\ref{fig:new_distribution_time_displaced}.\footnote{Of our sample, 9\% reported displacements that had not ended at the time of being surveyed.}

 \begin{figure}
\includegraphics[width=\linewidth]{graphics/Total_weights_q18_how_long_gone_in_total_overall.pdf}
\caption{Among displaced people, 42\% were displaced for less than seven days, 33\% for 7 to 14 days, 13\% for 14 to 30 days, and 9\% longer than a month.}
 \label{fig:new_distribution_time_displaced}       
\end{figure}

\subsection{Policy implications}

Of those displaced more than one night, 63.4\% report displacement of more than three nights: the typical displacement experience is a significant disruption to people's lives.  Furthermore, a disproportionate share of large families experience longer displacements.  These findings suggest household structure could be important for designing proper policy interventions.

Respondents did not report taking advantage of government-provided transportation\footnote{Though open to interpretation to respondents, this was intended to capture transportation or evacuation where a local, state, or national government provided transportation away from the home of a displaced person.  This transportation could take several forms (examples from past disasters are varied).  This refers to the question, "Did the government help you with transportation to leave your home?"  We could update to "government-assisted" transportation to be more accurate. However, this is perhaps a bit less clear.}, though we cannot establish the availability of such assistance reliably. Of those displaced, only 5.5\% of people reported using government-provided transportation.  Rather, most evacuated respondents reported using a car (89\%), which is likely influenced by local cultural factors.  The relative importance of private transportation in displacement has clear utility in informing policy responses for future similar disasters, and future surveys can demonstrate how local context and disaster type interact to produce various demands for transportation. This finding is also reflected in the behavioral data we have from aggregate trends in Facebook Displacement maps -- automated displacement estimation based on geolocated data -- used by nonprofits to allocate resources for numerous disasters worldwide.

We can also make inferences based on the patterns of displacement we observed.  The vast majority of displaced people (86\%) had returned home by the time they were surveyed (two months after the start of the fires).  Among displaced people, 42\% were displaced for less than seven days, 33\% for 7 to 14 days, 13\% for 14 to 30 days, and 9\% longer than a month. Those who had not returned home cited a range of reasons, including unsafe conditions at their original home (57.5\%), new opportunities in their post-displacement location (22.2\%), being otherwise unable or unwilling to return (9.7\%), or not being allowed to return (8\%).

\subsection{Future directions}

Surveys conducted on social media platforms such as Facebook provide a promising supplement or replacement for classical address-based samples in cases where populations are difficult or expensive to access (e.g., people displaced by a natural disaster). These methods can be further developed to provide rapid-response survey information on post-disaster conditions worldwide. Such a database of responses would provide a huge resource to demographers, policymakers, and researchers to help inform disaster responses. Further, work such as \citet{feehan2019using} provides a path forward to generalize these results from the Facebook population to the general population. Though more work needs to be done to understand the representativeness of this sampling frame, we feel this work is tractable and that the current analysis provides a promising first step to the advancement of our understanding of post-disaster displacement events.

\clearpage
\bibliographystyle{spbasic}
\bibliography{aufire.bib}
\clearpage

\appendix
\section*{Survey}

The survey can be found at \cite{maas2020using}.

\end{document}